\documentclass[12pt]{article}

\usepackage{scicite}
\usepackage{graphicx}
\usepackage{mathtools}

\usepackage{times}

\newenvironment{sciabstract}{%
\begin{quote} \bf}
{\end{quote}}

\newcounter{lastnote}

\title{{\it Quantum 3.0\/}: Quantum Learning, Quantum Heuristics and Beyond}

\author
{Mrittunjoy Guha Majumdar$^{1\ast}$\\
\\
\normalsize{$^{1}$Department of Physics, Amrita Vishwavidyapeetham,}\\
\normalsize{Nowlur, Amaravati, Andhra Pradesh 522503, India}\\
\\
\normalsize{$^\ast$To whom correspondence should be addressed; E-mail:  mrittunjoy@av.amrita.edu}
}

\date{}

\begin{document} 


\baselineskip24pt


\maketitle


\begin{sciabstract}
  Quantum learning paradigms address the question of how best to harness conceptual elements of quantum mechanics and information processing to improve operability and functionality of a computing system for specific tasks through experience. It is one of the fastest evolving framework, which lies at the intersection of physics, statistics and information processing, and is the next frontier for data sciences, machine learning and artificial intelligence. Progress in quantum learning paradigms is driven by multiple factors: need for more efficient data storage and computational speed, development of novel algorithms as well as structural resonances between specific physical systems and learning architectures. Given the demand for better computation methods for data-intensive processes in areas such as advanced scientific analysis and commerce as well as for facilitating more data-driven decision-making in education, energy, marketing, pharmaceuticals and health-care, finance and industry. 
\end{sciabstract}


\section*{Introduction}

\textit{Quantum 1.0} was the revolutionary utilization of quantum resources for technology, primarily electrical and optical, such as transistors and optical masers, while \textit{Quantum 2.0} was about harnessing non-classical elements in the quantum formalism such as entanglement for information processing \cite{bloembergen1964quantum, paoli1970observation, lehovec1949hole, hall1960tunnel, deutsch1985quantum, feynman1985quantum, peres1985reversible, zurek1984reversibility, benioff1980computer, divincenzo1995quantum, barenco1995elementary, shor1995scheme, divincenzo1995two, shor1994algorithms, cirac1995quantum}. Quantum mechanics have been used for undertaking information processing tasks such as teleportation, superdense coding, remote state preparation, quantum tomography, quantum cryptography and quantum key distribution, circuit-based and measurement-based quantum computing, quantum network coding and quantum internet, quantum random walks and quantum transduction \cite{bouwmeester1997experimental, furusawa1998unconditional, pirandola2015advances, guha2021nested, wang2005quantum, liu2002general, harrow2004superdense, bennett2001remote, dakic2012quantum, bennett2005remote, xia2007multiparty, jullien2014quantum, flammia2012quantum, gisin2002quantum, brassard2000limitations, scarani2009security, shor2000simple, gottesman2004security, steane1998quantum, o2007optical, preskill2018quantum, majumdar2018quantum, yurke1984quantum, hayashi2007quantum, simon2017towards, kimble2008quantum, wehner2018quantum, majumdar2021quantum, aharonov1993quantum, kempe2003quantum, travaglione2002implementing, shenvi2003quantum, majumdar2022polarization, lauk2020perspectives, rakher2010quantum, mirhosseini2020superconducting, wu2020microwave, andrews2014bidirectional, bagci2014optical, majumdar2023harnessing}. What can be called as \textit{Quantum 3.0} would be the harnessing of quantum resource and representation theory for a novel computational learning paradigm, along with quantum generalizations of existing classical computational learning models \cite{biamonte2017quantum, dunjko2018machine}. This is the next frontier of exploration in the quantum realm and seeks to address the question: can we fundamentally reframe and re-envision learning models when the information-system and/or process is quantum mechanical in nature?
\\
\\
The term 'machine learning' was coined by Arthur Samuel in 1959, in the context of a 'looking-ahead' algorithm implemented on a classical computer for a game of checkers, realized using a neural net approach with randomly connected switching net as well as an approach involving a highly organized network designed to learn only specific things \cite{samuel1959some}. In 1961, the punched-tape system known as Cybertron K-100 by Raytheon was developed as an early learning system that employed pattern recognition on sound samples, such as those from sonar signals, for learning \cite{scharff2017and}. Basic pattern recognition, storage and referencing were the primary elements that supported the learning framework in the next few decades, such as with the work on feedforward networks with one layer of modifiable weights connecting input units to output units, in what could be called reflexive systems that could discover hidden relationships in data \cite{nilsson1990mathematical, duda1973pattern}. Machine learning addresses the twin questions of what fundamental information theoretic laws govern all learning systems and how could one construct computing systems that can improve their functioning through experience \cite{jordan2015machine}. For various applications, training based on a limited set of inputs and outputs obtained from a system for specific empirical task seems to be more optimum than trying to predict the generalized response of the system in all circumstances. 
\\
\\
Learning, in the context of information theory, describes the improvement of a yardstick or performance criterion for a system for a given task. For instance, we could seek to classify Iris flowers into three species: \textit{Iris adriatica}, \textit{Iris taochia} and \textit{Iris kemaonensis} from a defining characteristic such as the cumulative length measurement of their petals and sepals. The yardstick index could be how accurately the classification takes place and the training could be over a historical sample of irises, from a phytologist's collection. We could also employ other criterion that have different penalties for incorrect classification. There are a number of algorithms for machine-learning that are applicable to different kinds of data-sets and problem-types. Machine learning algorithms probe through various candidate models, based on training data and experience, to find the one that is optimum for the selected performance metric. The diversity of machine-learning algorithms can be segregated based on their manner of representation of candidate models (such as decision trees) and their manner of probing through the space of candidate models (such as evolutionary search methods).
\\
\\
The question is whether loading our phytological data into quantum states can make this classification better. We could create a parameterized quantum circuit or \textit{ansatz} to do this, using a feature map. Any parameterized quantum circuit should be able to generate maximum number of quantum states in the Hilbert space and also have ability to entangle constituent qubits. It turns out that the variational quantum classifiers (VQCs) help classify the irises efficiently, using optimizers such as the constrained optimization by linear approximation optimizer (COBYLA). It is this interface of quantum mechanics and machine learning that provides various possibilities: machine learning with quantum computers, classical learning for quantum mechanical problems and generalized quantum learning theory. Quantum machine learning brings together the learning capability and scalable nature of machine learning and the speed, efficiency and processing power of quantum computers. There have been several classical learning paradigms that have had quantum analogues proposed and realized: quantum neural networks, quantum reinforcement learning, quantum support vector machines, quantum linear regression, cluster detection and assignment algorithms, quantum principal component analysis and quantum decision tree classifier \cite{menneer1995quantum, rebentrost2014quantum, schuld2016prediction, lu2014quantum, lloyd2014quantum, lloyd1982least, lloyd2013quantum, mckiernan2019automated}. In this comprehensive review of quantum learning paradigms, we will be looking at the spectrum of algorithms and protocols, before moving on to understanding the conceptual shift that this may provide when assessing learning theory from the perspective of quantum phenomenon. 
\section*{Quantum Patterns}
The underlying premise for the utility of the quantum realm for machine learning has to do with the emergence of classically anomalistic patterns in data. Among phenoma that are intrinsically quantum in nature, an interesting example is that of quantum revival patterns, where we find superposed and displaced replicas of the initial state of the system \cite{vasseur2015quantum, robinett2004quantum, e2023revival, berry2001quantum}. Quantum Kerr systems have such a kaleidoscopic mode of evolving, which is found to be reproducible based on interference of trajectories in the classical phase space \cite{lando2019quantum}. In the quantum-classical limit, where Planck's constant $\hbar\rightarrow 0$, it is seen that observables behave differently, leading to the preservation or disappearance of significant variables and their stochastic tendencies, such as for oscillator coordinate and spin variable respectively \cite{dumitru1995behavior}. Even in the case of skyrmions, we cannot simulate quantum skyrmions on classical supercomputers due to fundamental limitations, particularly around quantum fluctuations \cite{mazurenko2023estimating}. Quantum patterns are seen as periodic magnetic formations that are spontaneously formed in antiferromagnetic Bose-Einstein condensate \cite{kronjager2010spontaneous}. When ultracold atoms form quantum ferrofluids, we find a spontaneous development of coherent quantum density patterns leading to the formation of a super-solid \cite{hertkorn2021pattern}. When nanoscale lead films grown on silicon substrates are annealed to high temperatures, we observe intricate quantum patterns in the energy landscape \cite{czoschke2004quantum}. A beat pattern emerges in the quantum magnetoresistance of polar semiconductors without a centre of inversion symmetry, such as InAs/GaSb, when the carrier concentration is high \cite{rowe2001origin}. Patterns are ubiquitous in quantum chemistry, in properties such as chemical hardness and electronegativity \cite{v2011electronegativity}.
\\
\\
Given the ability of classical computers to both produce and recognize classical statistical data, the extrapolation would be that systems that produce classically atypical data, like in quantum systems, would also be able to recognize such data. This gives us the possibility of harnessing the quantum analogue of complex classical pattern recognition. Pattern recognition compares data input with specific memorized patterns, as it processes this input. Quantum pattern recognition provides a number of advantages over its classical counterpart. For instance, if we assume a Hopfield network approach for pattern recognition, we do so by local optimization, while utilising a quantum approach such as adiabatic quantum computation does so using global optimization \cite{szandala2015comparison}. A realization of this was using the liquid-state NMR technique, where some of the interesting insights included the possibility of representing a superposition of recognized patterns using a quantum neural register \cite{neigovzen2009quantum}. The parallelism inherent in quantum phenomena such as entanglement facilitate the execution of subroutines, even with big data, and this is one of the primary reasons for quantum machine learning, in general, being regarded as being better than its classical counterpart. Among pattern recognition protocols, one-class classification is important due to its applications in areas where detection of abnormal data points vis-à-vis instances of a known class is needed, using techniques such as support vector machines and principal component analysis \cite{tax2002one, khan2014one, khan2010survey, hoffmann2007kernel, scholkopf1999support, platt1999estimating}. This helps in addressing problems where we have imbalanced datasets, such as in metaheuristics, medical image datasets, manufacturing, high-energy physics and bioinformatics \cite{gharehchopogh2023quantum, gao2020handling, krawczyk2014weighted, du2019fault, lee2020fault, md2020alternate, spinosa2005combining, yousef2016feature}. Recently, a semi-supervised quantum one-class classification system known as \textit{Variational Quantum One-Class Classifier (VQOCC)} was developed using \cite{park2022variational}. At the centre of this proposition is a quantum autoencoder, which is basically a circuit that undertakes compression of a quantum state onto a lesser number of qubits while retaining the information encapsulated in the initial state \cite{romero2017quantum, pepper2019experimental}. Quantum autoencoders have been applied to quantum data compression, quantum error correction, denoising of data and even for preserving entanglement \cite{huang2020realization, locher2023quantum, bondarenko2020quantum, zhou2022preserving}. 
\\
\\
An important question to be addressed here is what can be categorized as a quantum pattern. Is it just an arbitrary pattern of quantum randomness? Is it the byproduct of a quantum process, such as in cold exciton gases? Is it the quantum analogue of classical patterns in nature? \textit{Harney} and \textit{Pirandola} defined a quantum pattern as an $m$-mode coherent state undergoing local $k$-ary modulations \cite{harney2022secure}. We can also define patterns in the information dynamics associated with quantum systems, such as in the case of quantum computing and communication systems \cite{leymann2019towards}. Examples would include basis encoding, quantum associative memory, amplitude encoding, angle encoding, QRAM encoding, quantum kernel estimator, variational quantum algorithm, variational quantum eigensolver, quantum associative memory, amplitude amplification, phase shift, oracle operations, quantum approximate optimization algorithm and quantum key exchange. Along with information theoretic arrangement and elements as well as quantum data types and quantum data structures, we can also specify entanglement structures to define the notion of quantum patterns \cite{perdrix2007quantum}. There are various kinds of entanglement patterns, from maximally entangled Dicke states to partially entangled cluster states. Even the outcome of measurements successively performed on an open quantum system have a pattern because of the interaction between system and environment. The pattern encapsulates information on non-Markovian memory effect and the relaxation rates associated with the system \cite{luchnikov2020machine}. The primacy of quantum patterns can also be seen within phases of matter, which can be distinguished by distinct symmetry breaking instances - within the Landau theory, symmetry of ordering of constituents of a physical system differentiates one phase from others. In certain systems, such as chiral spin liquids, even without symmetry breaking, we have different characteristics due to what is known as topological order, which has been posited to describe entanglement in many-body systems \cite{wen2019choreographed}. 
\section*{Quantum Algorithmic Resource-Pool}
The central question around assessing the possibilities in quantum machine learning is whether we have the algorithmic tools in the quantum domain for the same. A quantum algorithm is a succession of instructions for tackling a problem on a practical quantum computer \cite{montanaro2016quantum}. \textit{Cleve et al} highlighted that a common thread underlying all quantum algorithms can be ascertained when ``quantum computation is viewed as multiparticle interference" \cite{cleve1998quantum}. By and large, we have quantum search algorithms, quantum simulations of quantum systems on a quantum computer and algorithms premised on quantum implementations of the Fourier transform like Shor and Deutsch-Josza algorithm. Quantum supremacy or the enhancement of computational ability using quantum systems over that of classical counterparts has heralded the age of noisy intermediate-scale quantum (NISQ) technologies \cite{terhal2018quantum, preskill2018quantum}. One of the earliest quantum algorithms came with by formulation of an algorithmic solution to a special case of the hidden subgroup problem by Peter Shor \cite{shor1999polynomial}. Quantum Merlin-Arthur (QMA) completeness of a problem highlights that a supposed solution to it can be verified by a quantum computing system - a condition that has been extended to $k$ unentangled provers in QMA$(k)$ class problems \cite{jordan2010quantum, liu2007quantum}. When it comes to claims of quantum supremacy, we often speak of speedup, resource reduction, scaling and ability to address greater number as well as variety of problems. Quantum speedup can be contextually defined in terms of the asymptotic behavior of the ratio of the times taken by a specific classical and quantum algorithm for a particular problem when the size of the problem is made to be very large \cite{ronnow2014defining}. Speedup can be provable, like in Grover's algorithm, while in some cases the enhancement due to the quantum effects is not obvious, like in quantum annealing \cite{vedral2010elusive, crosson2021prospects}. Quantum resource optimization is seen in realization of randomness processing with quantum Bernoulli factories \cite{patel2019experimental}. 
\\
\\
The reason for enhancement of computational power, when it comes to machine learning and data analysis, with quantum machine learning arises due to the inherent primacy, within quantum mechanics, of high-dimensional vector spaces and, more importantly, matrix transformations between vectors in such spaces, as is important for data analytics and machine learning methods. The speedup is especially pronounced when the matrices being dealt with have an element of sparseness or are low-rank matrices \cite{harrow2009quantum, clader2013preconditioned}. A classic example arises in a key tool used for machine learning - \textit{principal component analysis}, which evaluates the eigensystem (and corresponding principal components) of the covariance matrix of data. Using a quantum random access memory, we initialize the vectors for a classical principal component analysis protocol onto quantum states and utilise the density matrix of the states (instead of the covariance matrix), which when exponentiated and operated on by a conditional SWAP operation for the quantum phase algorithm, yields the eigensystem that forms the premise of the principal component analysis \cite{lloyd2014quantum}. The algorithmic scaling is to the order of the square of the logarithm of the size of the system, in query as well as computational complexity. Essentially, the quantum principal component analysis scales exponentially faster than the classical counterpart.
\\
\\
While quantum principal component analysis is an unsupervised method that identifies patterns within higher-dimensional data for the reduction of data complexity with the retention of most of the information, there are various supervised learning algorithms that have had a quantum implementation \cite{schuld2018supervised}. One such algorithm relates to quantum support vectors. Support vector machines undertake regression and classification using the delineation of feature vectors of the data into distinct classes around a hyperplane that is at maximum distance from the nearest data-points in either classes around it \cite{hearst1998support}. If the data is not separable into distinct classes, we can use the kernel trick to project the data into a higher-dimensional space - a pursuit that by Cover's theorem can help in attaining separability in the data-set \cite{samuelson2011application}. Quantum algorithms such as Grover's algorithm are premised on binary classification. Even in cases of machine learning algorithms based on adiabatic quantum evolution, we define our approach in terms of a strong classifier that discerns whether a data-point is correct or erroneous based on a program specification \cite{pudenz2013quantum}. Like in the classical case, the quantum support vector machine relies on the definition of a quantum kernel that we can create by using a quantum feature map $\phi(\vec{x})$ between classical feature vectors $\vec{x}$ and a Hilbert space to help us obtain the kernel $K(\vec{x}_i,\vec{x}_j) = \vert\langle\phi(\vec{x}_i)\vert\phi(\vec{x}_j)\vert^2$. We can then expand the hyperplane in the general form: $f(\vec{x}) = \sum_i \alpha_i y_i K(\vec{x}_i,\vec{x})$, where $\alpha_i$ denote bounded positive quantities and $y_i$ the data labels. We can then use measures like the Gaussian or Rademacher Complexity to evaluate the classification error, based on this hyperplane definition and the data-set \cite{anguita2003quantum}. While the quantum support vector machine paradigm defined using Grover's algorithm gives a quadratic speedup, the implementation using the least squares approach provides an exponential speedup over classical algorithms \cite{rebentrost2014quantum}. In recent years, we have had quantum support vector machine implementations with the Newton method, amplitude estimation, gradient descent and using quantum annealers as well as variational quantum-circuitry \cite{zhang2022quantum, zhang2023quantum, li2022quantum, willsch2020support, ezawa2022variational}.
\begin{figure}[h]
    \centering
    \includegraphics[scale=0.45]{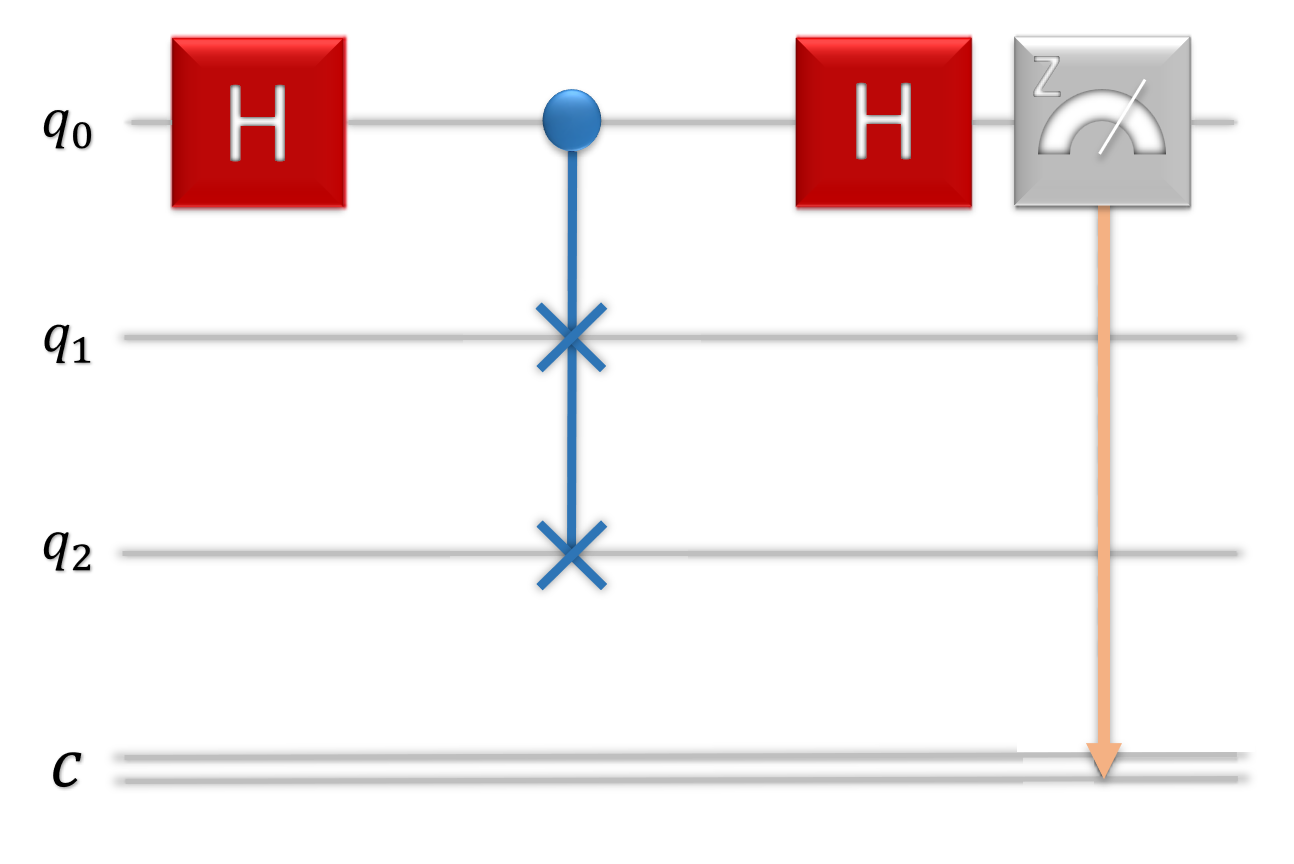}
    \caption{Swap Test}
    \label{fig:my_label}
\end{figure}
\\
\\
A natural extension of this comes with quantum $k$-nearest neighbour algorithms. Instead of a hyperplane as in the case of support vector machines, the underlying assumption of $k$-nearest neighbour methods is that the likelihood of two data-points that are proximal being of the same type is high \cite{cunningham2021k}. Labelled training vectors are used as reference for comparison of unlabelled testing vectors to determine $k$ nearest train-state neighbours for the specific testing vector-state, whose label is ascertained using majority voting. In the quantum picture, we use the concept of fidelity of a specific testing state $\vert\psi\rangle$ with respect to multiple training states $\vert\phi_j\rangle$: $F_j = \vert\langle\psi\vert\phi_j\rangle\vert^2$. After taking an initial set of candidate neighbour states, we use quantum search algorithms to find other states and corresponding indices till the nearest neighbours are found. The functional way to obtain the fidelity is by using the swap test. In this test (Figure 1), we begin with two quantum states $\vert a\rangle$ and $\vert b\rangle$ along with an ancilla qubit initialized to $\vert 0\rangle_{anc}$, on which we apply the Hadamard operation on the ancilla qubit. We thereafter apply a controlled swap operation on the ancilla qubit. A controlled swap operation is a gate where target qubits $\vert a\rangle$ and $\vert b\rangle$ are swapped if the control (in this case, the ancilla qubit) is in the state $\vert 1\rangle$. Applying a second Hadamard gate on the ancilla gives us the state $\vert\chi\rangle = \frac{1}{2}\vert0\rangle_{anc}(\vert a,b\rangle+\vert b,a\rangle)+\frac{1}{2}\vert1\rangle_{anc}(\vert a,b\rangle-\vert b,a\rangle)$. The measurement of the state $\vert0\rangle_{anc}$ has the associated probability $P(\vert0\rangle_{anc}) = \frac{1}{2}+\frac{1}{2}\vert\langle a\vert b\rangle\vert^2$. The inner product in this expression makes the overlap between the states primary: orthogonality gives us a probability of 0.5 while maximum overlaps gives a unity probability. This method can be used to find the distance between real-valued multi-dimensional vectors with the use of a quantum measurement \cite{lloyd2013quantum}. We can also undertake more rigorous pattern recognition between binary strings $\vert a_1, a_2,...,a_n\rangle$ and $\vert b_1, b_2,...,b_n\rangle$, using an extension of the construction in the $k$-nearest cluster model by initializing a state $\vert\psi\rangle = \vert a_1, a_2,...,a_n, b_1, b_2,...,b_n\rangle\otimes\frac{1}{\sqrt{2}}(\vert0\rangle+\vert1\rangle)_{anc}$, undertaking a XOR operation between the respective $(a_k, b_k) \forall k$ before storing it in place of the $b_k\forall k$, and finally undertaking a Hadamard operation on the ancilla qubit \cite{schuld2015introduction}. If we measure the ancilla qubit in the ground state $\vert0\rangle_{anc}$, we obtain a resultant state whose amplitude has a scaling characteristic dependent on the Hamming distance between the binary strings.
\section*{Quantum Reinforcement Learning and Deep Learning}
Going beyond supervised or unsupervised machine learning models, a major area of contemporary research has been in quantum-enhanced reinforcement learning, which is premised on the adaptive evolution of a quantum system based on reinforcement from a classical or quantum environment \cite{dong2008quantum}. The system receives \textit{percepts} from the environment and undertakes \textit{actions}. Unlike in conventional learning models, the learner in a reinforcement learning model has an influence on the state of the environment around it as much as it is influenced by it, thereby making it impossible to represent the environment in terms of a stationary memory. Both - the system and the environment, are stored as maps with memory, and the history of interactions between the two is the fundamental element in reinforcement learning. In the case of quantum reinforcement learning (QML), this history must be maintained in a quantum setting. In QML, we define a Hilbert space for the percept and action states - $\mathcal{H}_{\mathcal{S}}$ and $\mathcal{H}_{\mathcal{A}}$ respectively. The agent and environment both have memory registers ($R_A$ and $R_E$ respectively) to store the histories of the system-environment composite. We can model the interaction with a distinct Hilbert space $\mathcal{H}_{C}$ and we can characterize the agent (environment) my a series of \textit{Completely Positive Trace Preserving (CPTP)} maps $\{\mathcal{M}_i^{\mathcal{A}}\}_i$ ($\{\mathcal{M}_i^{\mathcal{E}}\}_i$) that acts on a resultant register formed by concatenation of the registers of system-interaction $R_AR_C$ (environment-interaction $R_CR_E$) systems. 
\begin{figure}
    \centering
    \includegraphics[scale=0.3]{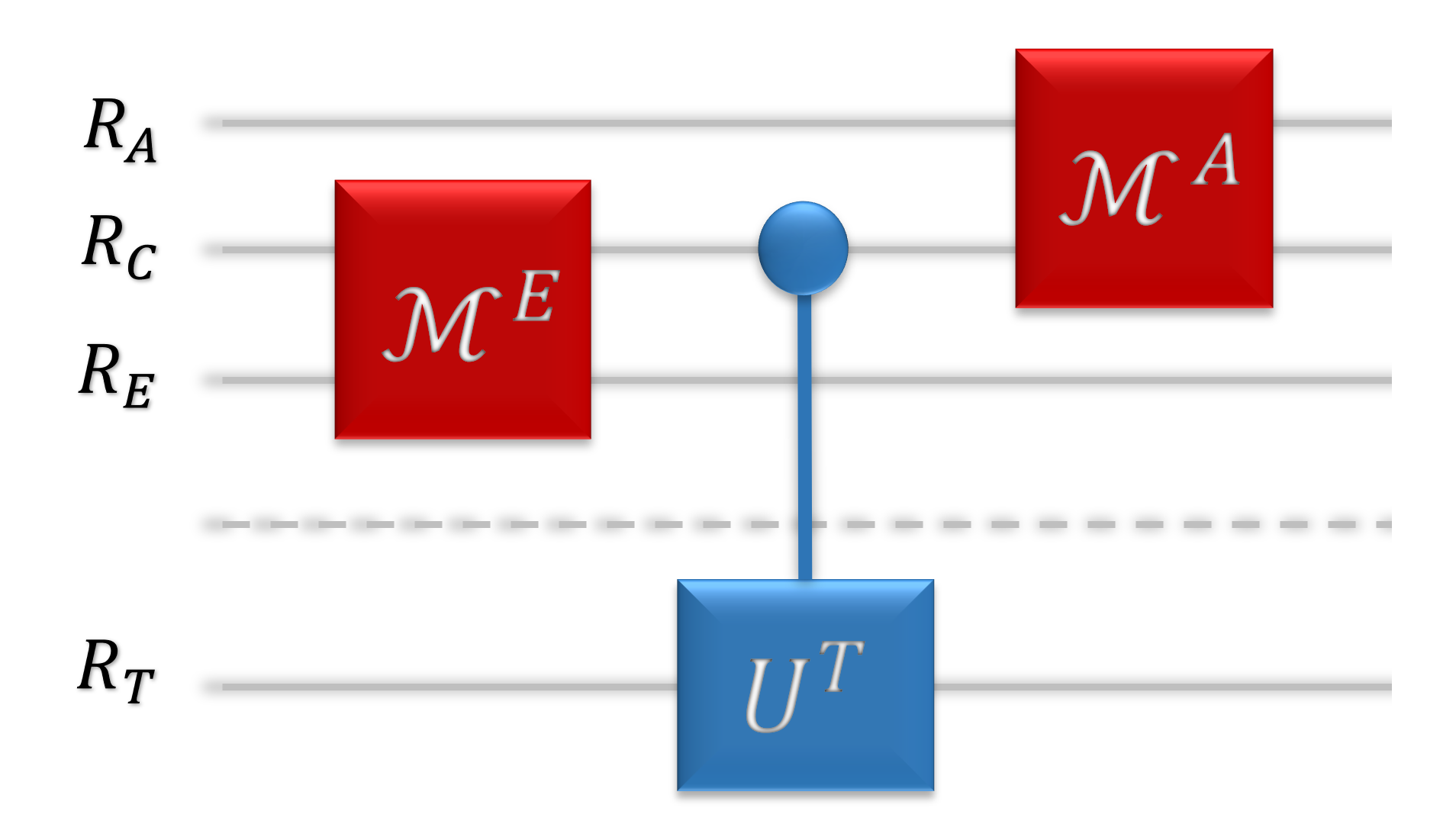}
    \caption{Quantum Reinforcement Learning (QRL) is premised on testing agent-environment interaction where the percept and action states have distinct Hilbert spaces associated with them. The register $R_T$ is not controlled by the agent or by the environment.}
    \label{fig:my_label}
\end{figure}
The performance of the system-environment is assessed against a figure of merit. \textit{Dunjko et al} showed that a quantum agent will outperform a classical learning agent associated with a classically delineated, controllable epochal environment against a particular figure of merit, if we were to consider a selected classical sporadic testing element \cite{dunjko2017advances}. The quantum enhancement arises from being able to extract additional attributes from the environment for optimization of a classical agent. Our point of interest is in environments that are quantum in nature and that facilitate the preservation of superposition of percepts and actions.
\\ 
\\
When we are talking of quantum reinforcement learning, it is imperative to discuss what we mean by quantum accessibility. Quantum accessibility implies the utilization of access of the agent to the environment for the simulation of an oracle $\mathcal{O}_q$ that behaves as follows: 
\begin{equation*}
\vert a_1, a_2, ..., a_k\rangle\vert x\rangle\xrightarrow{\mathcal{O}_q}\vert a_1, a_2, ..., a_k\rangle\vert x\oplus R(a_1,...,a_k)\rangle
\end{equation*}
where $R$ is the value of the 'reward' that the agent obtains once it successfully executes the actions $a_1, a_2,...,a_k$, for some auxiliary state $\vert x\rangle$. In the cumulative process, we see the utilization of a quantum interaction using what is known as \textit{oracularization} with the environment to undertake effective simulation of access to the oracle (which encapsulates important information about the environment), followed by the usage of this information for determining behavior that is optimum in the given environment. We can attain a greater insight into the oracularization process by understanding the import of an auxiliary state, which usually comprises of percept states $\vert s_1, s_2,...,s_k\rangle$, the reward state $\vert R\rangle$ and $\vert e_1,e_2,...,e_k\rangle$ that are auxiliary quantum subsystems kept within the environment. For the simulation of the oracle, these states need to be controlled and erased \cite{dunjko2017exponential}. To undertake this, the agent should be able to undertake modification of specific memory components of the environment with what are know as \textit{register scavenging} and \textit{register hijacking} operations, followed by a way to 'uncompute' with the implementation of the Hermitian adjoint of the net unitary (map) that the environment implements. There are two primary kinds of reinforcement learning paradigms with specific quantum realizations: \textit{value-based}, like in the case of Q-learning, which is premised on the learning of a value-function that guides how the system-environment makes decisions at each time step, as well as \textit{policy-gradient methods} that optimizes a policy (function) $\pi(a\vert s; \theta)$ using the parameter $\theta$ \cite{watkins1992q, skolik2022quantum, kakade2001natural, yao2020policy}. An important concept in this regard is meta-learning, wherein the learning agent undertakes the identification of its meta-parameter configuration that is optimum so as to be able to facilitate the optimization of performance. This parameter meta-learning is done by having the learning agent monitor its performance, and utilising a metaparameter register as well as techniques like adaptive Bayesian quantum estimation \cite{fiderer2021neural}. The reduction of the agent-environment interaction to a unitary oracular query is not feasible when we consider memory effects. In general, the oracular element may vary temporally and in such scenarios amplifying amplitude using Grover-type methods may be a possible approach to undertake reinforced learning in reward spaces that have an increasing monotonicity in the success probability. 
\\
\\
Deep learning has also seen a quantum expansion in recent years. Deep learning is premised on using artificial neural networks for discovering the representations required for the detection and classification of features from raw data \cite{lecun2015deep, goodfellow2016deep}. The basic unit of a neural network is a perceptron or a single artificial neuron. \textit{Gallant} described one of the first perceptron-based connectionist models where each cell $i$ computes a single activation $u_i$, which may be input to other cells or be an output of the network \cite{gallant1990perceptron}. Ricks et al gave one of the earliest quantum neural network models based on implementation of quantum circuitry with gates whose weights are evolved through learning using quantum search as well as piecewise weight allocation \cite{ricks2003training}. In each step, we have a density operator for the qubits representing the hidden states, which can can be extracted with any output ancilliaries using a partial trace operator and fed forward to the next layer of the neural network where the unitary transformations that encapsulate the action of perceptrons can be applied. There have been recent works on Bayesian approaches being implemented using quantum algorithms to learn Gaussian processes to train neural networks that are arbitrarily deep, with a quantum matrix inversion protocol being the core routine \cite{zhao2019bayesian}. In the quantum convolutional neural network presented by \textit{Cong et al} \cite{cong2019quantum}, the modeling of the convolutional layer is in terms of a unitary transformation (on the state density of the input) that is quasi-local and is applied in error correction and phase recognition in quantum systems. The forward pass of the convolutional neural network can be computed as a convolutional product using quantum algorithmic tools even as gradient descent methods can give us an idea about what are the relevant network parameters in the quantum context \cite{kerenidisquantum}. There have also been hybrid models with \textit{quanvolutional layers} in the network where we undertake local transformation of the data with several random quantum circuits in the set bounded-error quantum polynomial time \cite{henderson2020quanvolutional}. Recently, quanvolutional methods were used for physiological application such as for body part recognition using \textit{Hybrid Quantum Convolutional Neural Networks}, although in this case the classical counterpart was found to have a greater validation accuracy by 0.5\% \cite{gohel2022organ}.
\section*{Quantum Natural Language Processing, Quantum Advantage and the Path Forward}
Language is one of the earliest forms of representative evolution for humankind. Notwithstanding the tale of the Tower of Babel or the Gigantomachy myth, the diversity and complexity of language is of prime interest for computational representation and processing. Natural language processing is all about harnessing computational methods for the learning and production of content in human languages. We have come a long way from simply analysing the disparate linguistic forms and recognizing speech patterns to the creations of dialogues, translating speech from speech as well as the identification of emotional response of users towards services and products. The synchronic model of language aligns with the psycholinguistic understanding which highlights that language is dynamic \cite{barr1995synchronic}. In any speech-based natural language processing (NLP) system, we have computation based on phonetic, phonemic and prosodic rules, besides the morphemic analysis which is as relevant for other kinds of NLPs as well \cite{cho2016prosodic, yolchuyeva2020transformer, kak1987paninian}. Higher levels of natural language processing include the lexical, the syntactic, the semantic, the discourse-oriented and the pragmatic \cite{raskin1987linguistics}. When it comes to natural language processing, we can have quantitative statistical approaches like the Hidden Markov models, connectionist approaches that are based on inter-related fundamental processing units as well as symbolic approaches that lay emphasis on the logic or rules that a linguistic framework encapsulates. \textit{Zeng} and \textit{Coecke} were among the first to give us a quantum version of natural language processing when they employed the closest vector problem for sentence similarity identification within the \textit{distributional
compositional categorical} (DisCoCat) model given by \textit{Coecke}, \textit{Sadrzadeh} and \textit{Clark} \cite{zeng2016quantum, wiebe2015quantum, clark2008compositional}. In the compositional distribution model, meanings are encapsulated in quantum states while quantum measurements give us the grammatical structure for a specific sentence and context \cite{guarasci2022quantum}. The embedding of the language in the vector space of quantum systems naturally leads to word-correlations represented by the vector geometry, even as we map digrams formed from parsing of sentences to quantum variational circuits. Essential quantum properties such as superposition helps us in the modelling of uncertainties in the language while phenomena like entanglement help us in describing how semantics and syntax are composed and distributed. Recently, natural language processing experiments have been practically realized for over 100 sentences on the \textit{Noisy Intermediate-Scale Quantum (NISQ)} computing system provided by IBM Quantum \cite{lorenz2023qnlp}. 
\\
\\
When we speak about quantum natural language processing, in particular, and machine learning, in general, the elephant in the room is: how much of an advantage do we obtain using quantum systems over classical ones? The 'quantum native' view of quantum natural language processing which posits that the vector-landscape provided by quantum systems is better suited for the linguistic - syntactic, semantic and pragmatic elements, is preliminary and unsubstantiated by practical realizations, with a critical shortcoming being the primary reliance on quantum RAMs, which are expensive and not yet empirically implemented. More broadly, there are many aspects of quantum computing that are often off-set by machine learning paradigms. For instance, the size of inputs is usually small, such as in fault-tolerant algorithms, while there could high-dimensional tensorial structures with numerous entries in machine learning. The problems in machine learning can be highly unstructured and complex while problems studied in quantum computing are structured, often with elements of regularity and periodicity. Theory is defined and delineated in quantum computing problems while empiricism is preeminent in machine learning, with modelling and interpolations being important in the latter. Quantum computing can have absolute benchmarks like the scaling of run-time while machine learning have more constructivist benchmarks between disparate models. Quantum advantage can be spoken of in terms of solvability, expressivity of the class of the model, size of training sample needed, generalizability and how the optimization landscape is structured \cite{liu2021rigorous, liu2023quantum, meyer2023exploiting, abedi2023quantum, arunachalam2018optimal, banchi2021generalization, krunic2022quantum, ahmad2021quantum, mcclean2018barren, ge2021optimization}. Due to the need to be able to capture the system dynamics in terms of quantum systems and circuitry, we can only gauge performance of machine learning algorithms in specific selected contexts and examples, with this selectivity preventing any generalizability of performance-based advantage, if any. Such constraints also exist for readout and loading of data, when it comes to quantum examples. Performance parameters are also highly contextual and it is not straightforward to assess if any 'quantum advantage' is due to the specific way in which we have selected benchmarking threshholds and hyper-parameters or whether they are actually observations made structurally. There may be a need to move towards studying \textit{Quantum 3.0}, the revolution of quantum learning theory, as a paradigm in itself, without needing to resort to bylines towards 'quantum advantage' and 'quantum supremacy'. For example, classical intractability of kernels being bypassed using quantum tools can be an interesting pursuit but given that such kernels are not found to be utilizable in realizable machine learning, such an approach has no practical relevance. Today, we have moved beyond kernel methods with even more efficient ones like data re-uploading and explicit models \cite{perez2020data, havlivcek2019supervised}. We now have quantum machine learning models that have features that preserve privacy as well as those that take into consideration the impact of decoherence \cite{watkins2023quantum, liao2023decohering}. The future is bright for quantum machine learning, even as the intersection of quantum mechanics, learning theory and information processing is explored to find novel methods of deploying artificial intelligence. 

\bibliography{scibib}

\bibliographystyle{Science}



\end{document}